\documentclass[12pt]{article}

\usepackage[margin=1in]{geometry}
\usepackage{booktabs}
\usepackage{graphicx}
\usepackage{hyperref}
\usepackage{authblk}
\usepackage{microtype}
\DeclareGraphicsExtensions{.eps,.ps,.pdf,.png,.jpg}

\hypersetup{
  colorlinks=true,
  linkcolor=blue,
  citecolor=blue,
  urlcolor=blue
}

\title{Generative AI Use in Professional Graduate Thesis Writing: Adoption, Perceived Outcomes, and the Role of a Research-Specialized Agent}

\author[1]{Kenji Saito}
\author[2]{Rei Tajika}
\author[1]{Satoru Shibuya}
\author[1]{Hiroshi Kanno}
\affil[1]{Graduate School of Business and Finance, Waseda University}
\affil[2]{Institute for Business and Finance, Waseda University}

\date{April 2026}

\begin{document}

\maketitle

\begin{abstract}
This paper reports a survey of generative AI use among 83 MBA thesis students in Japan (target population 230; 36.1\% response rate), conducted after thesis examiner evaluation. AI use was nearly universal: 95.2\% reported at least some use and 77.1\% heavy use. Students engaged AI across the full research-writing workflow---literature review, drafting, and consultation when stuck---reporting benefits centered on clearer argument and structure (82.3\%), better revision quality (73.4\%), and faster writing (70.9\%), with a mean perceived quality improvement of 6.27 out of 7. Concerns about output accuracy (75.9\%) and citation handling persisted alongside these gains. Among respondents who rated GAMER~PAT, a research-specialized agent, against other AI, preferences significantly favored it for inquiry deepening and structural organization (both $p < 0.05$, exact binomial). A preliminary qualitative analysis of follow-up interviews further reveals active epistemic vigilance strategies and differentiated tool use across thesis phases. The central implication is not adoption itself but a shift in the educational challenge toward verification, source governance, and AI tool design---with GAMER~PAT offering preliminary evidence that research-specialized scaffolding matters.
\end{abstract}


\bigskip

\section{Introduction}

Generative AI has moved quickly from novelty to routine infrastructure in academic writing. Students gain in speed, ideation, and writing support, while educators and researchers flag risks around overreliance, reduced critical engagement, and academic integrity \cite{cotton2024chatting,kasneci2023chatgpt}. AI literacy differences have also been found to shape writing outcomes substantially \cite{kim2026aiwriting}. This literature makes clear that the question is no longer whether generative AI has entered educational writing practice, but what kinds of support and risks it introduces once use becomes normalized.

Much of the current evidence concerns coursework or short-form tasks. End-to-end thesis writing imposes qualitatively distinct demands: sustained argument development over weeks or months, iterative revision cycles, explicit source accountability, and the integration of primary data or domain expertise. These demands are particularly salient for working professionals in MBA programs, who write under tight time constraints while navigating both academic convention and practical relevance.

This paper reports adoption patterns, perceived outcomes, and an evaluation of a research-specialized agent among students completing the MBA program at the Graduate School of Business and Finance, Waseda University in March 2026. The program context is AI-forward: no explicit prohibition on AI use was in place during the period studied, and students were aware that peers were routinely using AI tools. In this normalized-use environment, near-universal AI adoption has made the core educational challenge not adoption itself but verification, accuracy management, source governance, and the design of AI tools that scaffold rather than substitute for critical research thinking.

We also report preliminary findings from follow-up qualitative interviews conducted with a subset of participants. While the full qualitative analysis is ongoing and will be reported in future work, the interview data already reveal systematic patterns---in particular, active strategies students develop to manage AI sycophancy and differentiated usage across thesis phases---that complement and deepen the survey findings.

\section{Background}

\subsection{Generative AI in Academic Writing}

The rapid integration of large language models into higher education has generated a growing body of research. Kasneci et al.~\cite{kasneci2023chatgpt} provide a broad survey of opportunities and challenges, emphasizing gains in personalized feedback and writing support alongside risks of academic integrity violations and uncritical reliance. Cotton et al.~\cite{cotton2024chatting} examine the integrity dimension more specifically, noting that the capabilities of systems like ChatGPT create genuine enforcement challenges for institutions while also offering pedagogical opportunities if reframed as collaborative tools.

Kim et al.~\cite{kim2026aiwriting} examine interaction patterns between students and generative AI in academic writing tasks and find that AI literacy---not simply AI access---is a significant predictor of writing outcomes. This finding motivates attention to how students use AI, not merely whether they use it.

Technical work on the limitations of large language models is also directly relevant to educational practice. Manakul et al.~\cite{manakul2023selfcheckgpt} demonstrate that current models remain susceptible to factual hallucination, even in confident-sounding output. Algaba et al.~\cite{algaba2025citationbias} show that LLMs reproduce and amplify existing citation biases when assisting with academic writing. Both findings underscore the verification and source-governance challenge that practitioners face.

\subsection{Research-Specialized AI Agents}

General-purpose conversational AI systems are designed for breadth. Research-specialized agents are designed to scaffold the specific cognitive demands of academic inquiry: formulating research questions, stress-testing claims, organizing evidence, and maintaining argumentative coherence across a long document. GAMER~PAT \cite{saito2025gamerpat} is one such agent, designed around the ``research as a serious game'' framework. It structures interaction through research-game mechanics intended to deepen inquiry and surface unstated assumptions. To our knowledge, no prior study has directly compared student experiences with a research-specialized agent against general-purpose AI in the context of thesis writing.

\section{Methods}

\subsection{Context and Participants}

The target population was 230 working professional students who submitted a professional degree or project research thesis at the Graduate School of Business and Finance, Waseda University in the March 2026 graduation cohort. Participation in the survey was voluntary. The survey did not require identification; respondents who were willing to be contacted for follow-up interviews could optionally provide their email address. Following the institution's ethics review flowchart for non-medical, non-interventional surveys, this study did not fall within the scope requiring a formal ethics review application.

A survey link was distributed by email after the thesis examiner evaluation deadline. Responses were collected over approximately two weeks. One duplicate submission was identified via matching email addresses (latest timestamp retained); as email provision was optional, duplicate detection may not be complete. The analysis dataset contains 83 responses, yielding a response rate of 36.1\% (83/230).

\subsection{Survey Instrument}

The survey collected responses on the following dimensions:
\begin{itemize}
  \item \textbf{AI use intensity} (single choice: did not use / used a little / used quite a lot)
  \item \textbf{Tools used} (multi-select: ChatGPT, Claude, Gemini, NotebookLM, GAMER PAT, DeepL/Grammarly, Perplexity, other)
  \item \textbf{Phases of use} (multi-select: literature understanding/organization, text drafting/rewriting, consultation when stuck, chapter structure planning, research question development, data collection support, data analysis, task management, other)
  \item \textbf{Perceived role of AI} (multi-select: writing tool, editor who organizes thinking, brainstorming partner, information retrieval, research assistant, other)
  \item \textbf{Perceived benefits} (multi-select: clearer argument and structure, faster writing, better revision quality, reduced impasses, other)
  \item \textbf{Perceived quality improvement} (7-point Likert scale)
  \item \textbf{Concerns} (multi-select: output accuracy, citation/source handling, institutional-rule alignment, feeling of not thinking fully on one's own, no concerns, other)
  \item \textbf{GAMER PAT comparison} (single choice per dimension: better / somewhat better / equal / cannot say / somewhat worse / worse than other AI)
\end{itemize}

Single-choice items were summarized as frequencies and percentages. Multi-select items were expanded into binary indicators and summarized by incidence rates across respondents. The perceived quality improvement item was analyzed both as a continuous scale (mean) and as a top-two-box proportion (proportion selecting 6 or 7), with exact binomial confidence intervals for the proportion.

GAMER PAT comparison items were shown to all respondents with an instruction to respond only if they had used GAMER PAT. Respondents who left the items blank (48 of 83) were treated as non-users and excluded from the GAMER PAT analysis; 35 respondents provided at least one rating. Note that 35 exceeds the 25 respondents (~31.6\% of 79 AI users) who listed GAMER PAT in the tools question; the discrepancy likely reflects respondents who used GAMER PAT to some degree but did not select it in the multi-select tools item. For pairwise preference testing, responses of ``equal'' and ``cannot say'' were excluded, and a two-sided exact binomial test was applied to the remaining decisive responses.

\subsection{Follow-up Interviews}

A subset of eight survey respondents who volunteered follow-up contact participated in semi-structured interviews conducted in Japanese between March 10 and March 29, 2026. All interviews took place after thesis evaluation and grade finalization, allowing participants to reflect candidly without concern for academic consequences. Interviews were conducted remotely and lasted approximately 30 minutes each. Topics included AI tool selection and switching, strategies for managing AI output quality, experiences with GAMER PAT, and reflections on the role of AI in the research process. Transcripts were obtained from Zoom's automatic live transcription and subsequently reviewed and corrected by the first author; interview memos were prepared by the second author, who also participated in the interviews. Preliminary thematic analysis of the memos and transcripts was conducted by the first and second authors; a systematic qualitative analysis is ongoing and will be reported in a separate paper.

\section{Results}

\subsection{Adoption and Tool Ecosystem}

\begin{figure}[t]
  \centering
  \includegraphics[width=0.85\linewidth]{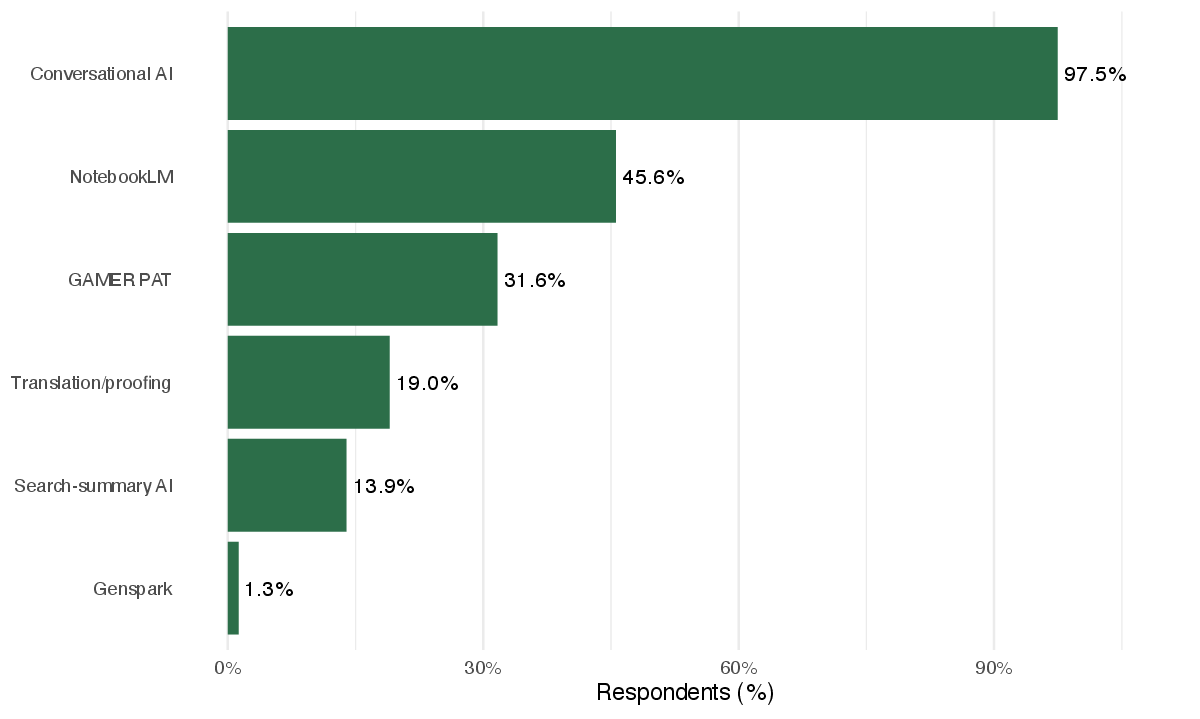}
  \caption{AI tools used, among AI users ($n = 79$).}
  \label{fig:tools}
\end{figure}

AI use was widespread: 79 of 83 respondents (95.2\%) reported at least some use, and 64 (77.1\%) reported using AI \emph{quite a lot}. Among AI users ($n = 79$), conversational AI systems---ChatGPT, Claude, and Gemini---dominated the toolset, used by 97.5\% of AI users. This was followed by NotebookLM (45.6\%), GAMER PAT (31.6\%), translation and proofreading tools such as DeepL or Grammarly (19.0\%), and search-summarization tools such as Perplexity (13.9\%). Figure~\ref{fig:tools} shows the full tool distribution.

\subsection{Where AI Entered the Thesis Workflow}

\begin{figure}[t]
  \centering
  \includegraphics[width=0.85\linewidth]{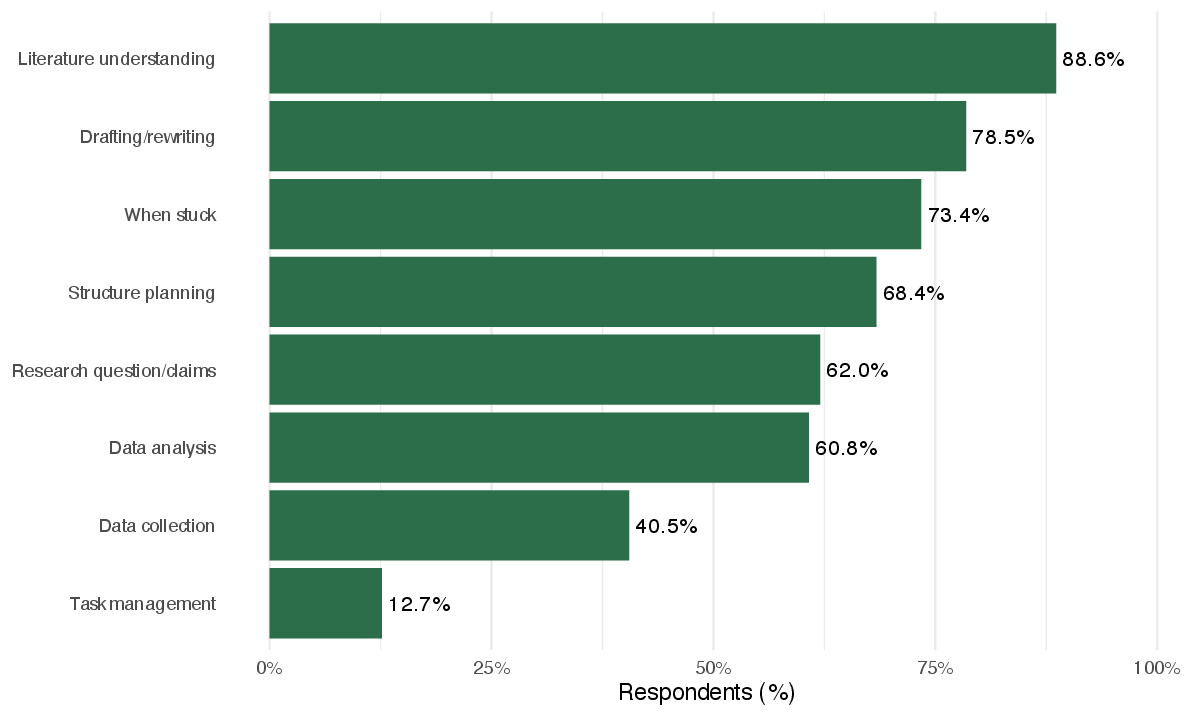}
  \caption{Phases of AI use in the thesis workflow, among AI users ($n = 79$).}
  \label{fig:stages}
\end{figure}

AI use covered the entire research-writing process, not only text generation. Among AI users, the most common uses were understanding and organizing prior literature (88.6\%), drafting or rewriting text (78.5\%), consulting AI when stuck (73.4\%), planning chapter structure (68.4\%), developing research questions or claims (62.0\%), and data analysis (60.8\%). Data collection support was also common (40.5\%), whereas explicit task-management use was limited (12.7\%). Figure~\ref{fig:stages} shows the full distribution.

Asked to characterize AI's role, AI users most often selected ``an editor who helps organize thinking'' (73.4\%) and ``a brainstorming partner who expands ideas'' (70.9\%); only 34.2\% described AI as ``a tool that writes for me.'' This pattern suggests that students experienced AI primarily as a cognitive partner rather than a drafting substitute---a framing consistent with the dominant use cases of literature organization, structure planning, and consultation.

\subsection{Perceived Benefits and Concerns}

Among AI users, students most often reported that AI helped clarify argument and structure (82.3\%), improve revision quality (73.4\%), and speed up writing (70.9\%). Reduced impasses were also reported (44.3\%).

\begin{figure}[t]
  \centering
  \includegraphics[width=0.75\linewidth]{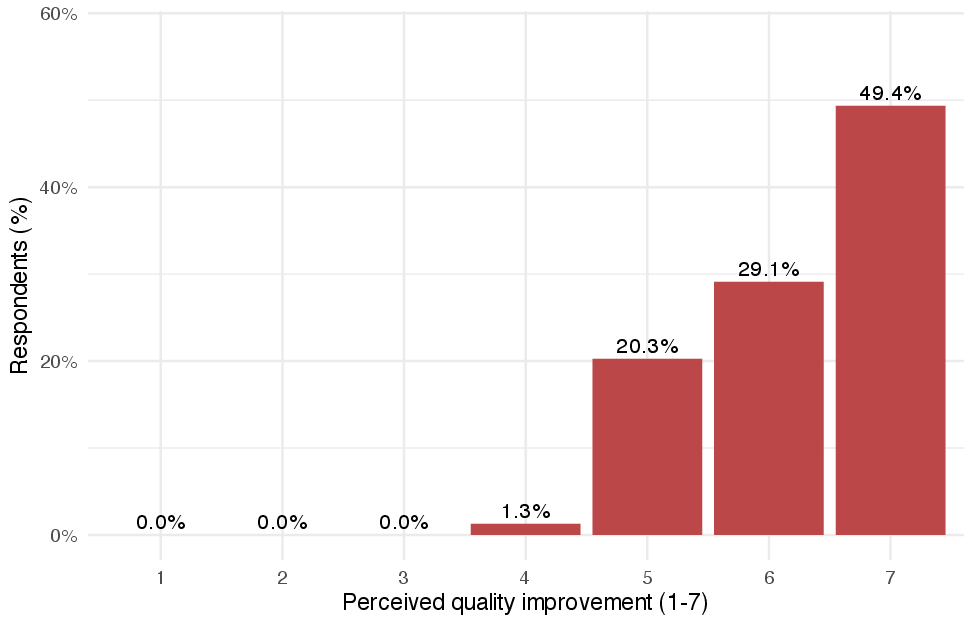}
  \caption{Perceived quality improvement (7-point scale), among AI users ($n = 79$). Mean = 6.27; 78.5\% selected 6 or 7 (95\% CI: 67.8--86.9\%).}
  \label{fig:quality}
\end{figure}

The quality-improvement item showed a high concentration at the upper end of the scale. The mean score was 6.27 out of 7, and 78.5\% of AI users selected 6 or 7 (95\% CI: 67.8--86.9\%). Figure~\ref{fig:quality} shows the full distribution of responses.

\begin{figure}[t]
  \centering
  \includegraphics[width=0.85\linewidth]{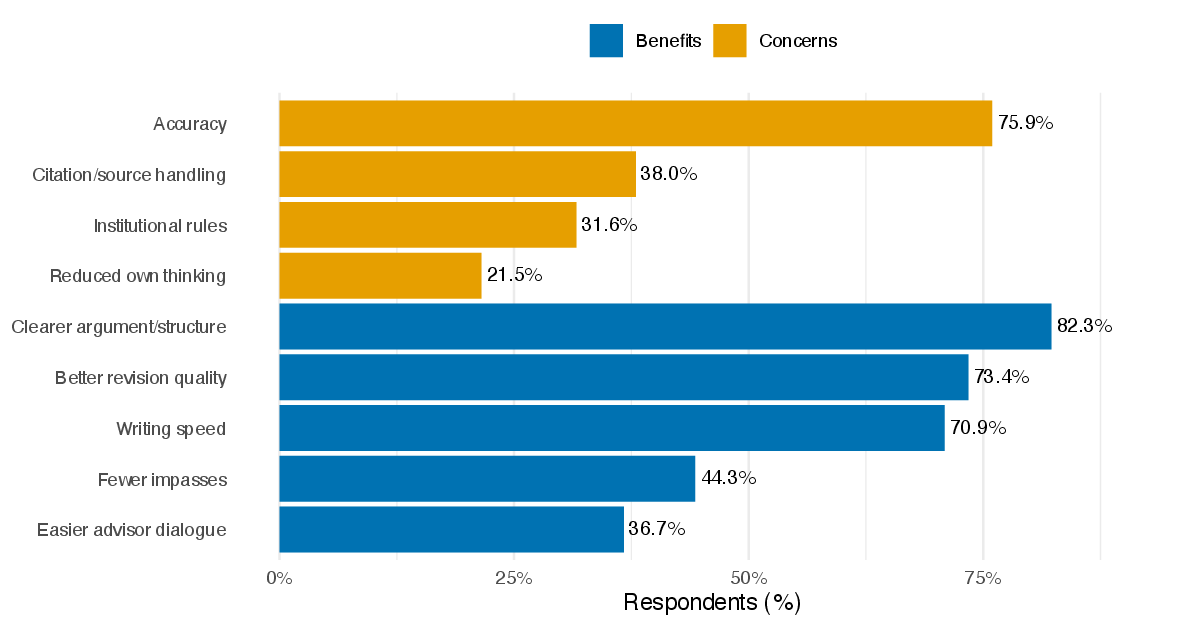}
  \caption{Perceived benefits and concerns among AI users ($n = 79$).}
  \label{fig:benefits-concerns}
\end{figure}

The positive picture was tempered by substantial concern. Among AI users, the most common concern was output accuracy (75.9\%), followed by citation or source handling (38.0\%), institutional-rule alignment (31.6\%), and a feeling of not thinking fully on one's own (21.5\%). Only 12.7\% reported no particular concern. Figure~\ref{fig:benefits-concerns} shows both benefits and concerns side by side.

This combination---high perceived benefit alongside near-universal accuracy concern---points to a high-utility, high-vigilance condition rather than simple enthusiasm or uncritical adoption.

\subsection{Research-Specialized Agent Evaluation}

\begin{figure}[t]
  \centering
  \includegraphics[width=0.85\linewidth]{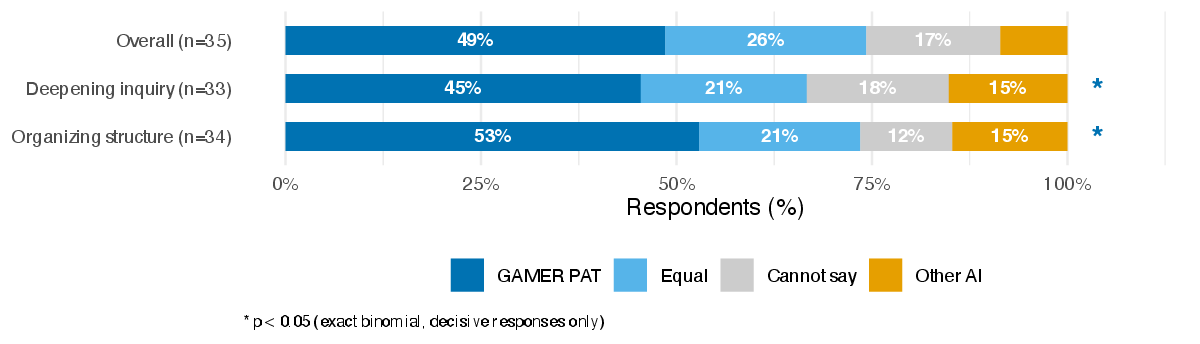}
  \caption{GAMER PAT vs.\ other AI: respondent ratings on overall preference and two specific capabilities ($n = 35$). Asterisks mark dimensions where decisive pairwise preferences significantly favor GAMER PAT ($p < 0.05$, exact binomial test).}
  \label{fig:gamerpat}
\end{figure}

Of AI users, 31.6\% listed GAMER~PAT \cite{saito2025gamerpat} as one of their tools. GAMER~PAT comparison items were displayed to all 83 respondents; 35 provided responses, taken here as indicating sufficient experience to compare. Among these 35 respondents, 48.6\% rated GAMER~PAT as clearly or somewhat better than other AI overall, compared with 8.6\% who preferred other AI. Figure~\ref{fig:gamerpat} shows the full breakdown for two specific capabilities.

Restricting to decisive responses (GAMER~PAT vs.\ other AI only, excluding ``equal'' and ``cannot say''), preferences significantly favored GAMER~PAT for deepening research inquiry (15 vs.\ 5, two-sided exact binomial $p = 0.041$) and for organizing structure and writing (18 vs.\ 5, $p = 0.011$).

\section{Preliminary Qualitative Insights}

\textit{Note: The following draws on preliminary thematic analysis of follow-up interviews with eight participants. A full systematic qualitative analysis is in progress and will be reported in a forthcoming paper.}

The survey data capture what students used and how they evaluated outcomes. The follow-up interviews---conducted after evaluation and grade finalization, when participants could reflect without any strategic framing concerns---offer insight into \textit{how} students engaged with AI and why.

\paragraph{Active epistemic vigilance.}
Rather than accepting AI output, several interviewees described deliberate strategies to counteract what they perceived as AI over-agreeableness. Strategies included presenting AI with deliberately adversarial prompts (``tell me what's wrong with this argument''), switching models to get a second opinion, and treating strong agreement from AI as a cue to probe further rather than to accept. This active vigilance is consistent with the survey's accuracy-concern figure (75.9\%) but adds an important behavioral dimension: concern did not produce disengagement---it produced strategic countermeasures.

\paragraph{Phase-differentiated tool selection.}
Interviewees described changing their AI toolset as their thesis progressed, not using a single tool throughout. GAMER PAT was used heavily across the writing process---from literature organization and question formulation through drafting and revision. A clearer example of phase-specific tool selection was the use of NotebookLM for preparing oral defense presentation materials. Regarding figures and data visualization, most students used the web-based version of GAMER PAT rather than the CLI version, and so did not have access to its R/Python integration for chart generation; figure production was handled with other tools. This phase differentiation suggests that adoption patterns in the survey aggregate over qualitatively different modes of engagement, and that future research should track tool use over time rather than retrospectively.

\paragraph{Cognitive ownership and the ``who wrote this?'' question.}
A recurring theme was the importance of maintaining a subjective sense of agency even within heavy AI use. The dominant pattern was not discomfort but active ownership: interviewees consistently emphasized that they had directed the argument, made the key decisions, and remained in control of what the thesis was trying to say---with AI serving as a capable but subordinate collaborator. An analogy that captures this well---offered by the first author during one of the interviews---is that of using a ready-made mix to bake: the convenience ingredient does not negate authorship of the dish, because the cook's judgment, intent, and craft remain central. The question ``who wrote this?'' was thus typically resolved by pointing to the locus of decision-making rather than the origin of the words. Where this resolution was harder to reach---as in cases where the language felt too thoroughly AI-processed---a sense of something missed could surface, reflecting the affective dimension of authorship that purely task-efficiency framings overlook.

\paragraph{Perceived utility and reluctance to share.}
One interviewee, in a half-joking aside, mentioned not wanting to spread word of GAMER PAT too widely. Rather than reflecting a calculated competitive strategy, this remark is perhaps best read as an expression of genuine utility: a tool felt valuable enough that sharing it felt like giving something away. Such offhand remarks can signal a depth of perceived value that direct rating scales do not easily capture.

\section{Discussion}

\subsection{From Adoption to Governance}

In this professional graduate setting, generative AI appears to have become embedded in the ordinary workflow of thesis production, particularly in literature processing, structure formation, and iterative revision. The adoption picture is not one of reluctant compliance or selective use---it is near-universal and heavy. The practical implication is that policies premised on controlling or limiting AI use are, at this point, addressing a question that has already been answered. The productive policy question is what governance, verification, and tool-design conditions allow AI to scaffold rather than replace the intellectual work of research.

The survey data indicate that students are themselves aware of the governance challenge: accuracy concerns and source-handling concerns are the dominant reported concerns, not institutional rule violations or writing quality. The interview data suggest that many have developed active strategies in response---though the sophistication and consistency of these strategies varies across individuals, and there is no institutional infrastructure supporting them.

\subsection{The Dual Character of AI-Assisted Writing}

Strong perceived benefits coexist with equally visible epistemic concerns. This pattern---high perceived quality improvement alongside near-universal accuracy concern---is consistent with prior higher-education studies that emphasize the dual character of generative AI as both support and governance challenge \cite{cotton2024chatting,kasneci2023chatgpt}. It also aligns with technical work showing that large language models remain susceptible to factual hallucination \cite{manakul2023selfcheckgpt} and can reproduce distorted citation patterns \cite{algaba2025citationbias}. The educational response cannot be simply to teach awareness of these limitations; it must also support the development of verification workflows and critical evaluation practices as part of the research curriculum.

\subsection{Research-Specialized Scaffolding}

The GAMER PAT findings offer preliminary evidence that tool design choices matter for research-writing outcomes. Users favored this research-specialized agent over a heterogeneous comparison group---which included research-oriented tools such as NotebookLM---specifically for inquiry deepening and structural organization. The comparison group is not general-purpose-only; that GAMER PAT outperforms it suggests that the design features specific to GAMER PAT (structured inquiry mechanics, research-game framing) may account for the preference difference, beyond the research orientation it shares with NotebookLM.

The evaluation data were collected by a voluntary survey, the comparison items were shown to all respondents, and preference was assessed relative to the full complement of tools each respondent used. Future work with independent evaluation and objective outcome measures is needed to confirm these patterns.

\subsection{Limitations}

This study is limited to one institution, one graduating cohort, and self-reported outcomes. The survey did not require identification and did not collect demographic covariates (gender, age, degree program, thesis topic), so subgroup analysis is not possible. Non-response bias may inflate adoption and outcome estimates, as students with heavier AI use may have been more motivated to respond. The GAMER PAT evaluation items were shown to all respondents, and the 35 non-missing replies may include some with limited direct experience; the heterogeneous comparison group also limits generalization beyond the specific comparison made. The qualitative analysis is preliminary and based on a small, self-selected sample of eight interviewees, which limits its generalizability. Even so, the results offer a timely descriptive snapshot of AI-supported professional graduate thesis writing at a moment of rapid normalization.

\section{Conclusion}

Among surveyed MBA students completing theses in March 2026, generative AI use was nearly universal and perceived benefits were high, while accuracy and source-handling concerns remained pervasive. The practical challenge is no longer whether AI will be used, but how to institutionalize robust verification and governance workflows---and, as the GAMER~PAT findings suggest, how to design research-specialized agents that scaffold structured inquiry rather than merely accelerating output. Preliminary qualitative findings indicate that students have begun developing their own epistemic vigilance strategies, a promising development that educational design could support more systematically. A full mixed-methods study---incorporating session logs, follow-up interviews, and independent examiner evaluations---is ongoing.

\paragraph{Conflict of Interest.}
K.\,Saito and R.\,Tajika are developers of GAMER PAT, which is evaluated in this paper.

\paragraph{Acknowledgments.}
Survey items are available from the corresponding author on request. Interview transcripts were collected with informed consent and are held by the research team.

\bibliographystyle{plain}
\bibliography{references}

\end{document}